# Generating hot carriers in plasmonic nanoparticles: when quantization does and does not matter?

Jacob B Khurgin [(1)], Uriel Levy [(2)]

[(1)] Department of Electrical and Computer Engineering, Johns Hopkins University, Baltimore, Maryland 21218, United States

[(2)] Department of Applied Physics, The Benin School of Engineering and Computer Science, The Center for Nanoscience and Nanotechnology, The Hebrew University of Jerusalem, Jerusalem, 91904, Israel

**Abstract:** Plasmon-assisted hot carrier processes in metal nanoparticles can be described either classically or using the full strength of quantum mechanics. We reconfirm that from the practical applications point of view, when it comes to description of the decay of plasmons in nanoparticles, classical description is sufficiently adequate for all but the smallest of the nanoparticles. At the same time, the electron temperature rise in nanoparticles is discrete (quantized) and neglecting this fact can lead to significant underestimating of hot carrier assisted effects, such as photocatalysis.

**Introduction**

The field of plasmonics[1, 2], active since early 21[st] century, have seen most recently two extremely interesting developments, both prompted by the realization that due to the inherently high losses [3] occurring in the metal, plasmonic devices will always remain at a disadvantage when it comes to many "conventional" nanophotonic devices, be that interconnects, sources, modulators, and so on. Therefore, it has become imperative to find a niche (or niches) where plasmonics will be competitive with existing dielectric-based nanophotonics. One such niche involves using so-called hot carriers generated by plasmon decay [4, 5] for photo-detection [6], energy harvesting[7], or for initiation of chemical reactions [8, 9]. In these applications, the strong loss in metal, which is typically thought of as the major drawback of plasmonics, is turned into its strength. The other niche where plasmonics may thrive and where activity is rapidly growing is a broadly-defined area of "quantum plasmonics"[10-12]. The operational principle of quantum plasmonic devices involves single quanta (photons or plasmons), and the relatively large loss can be tolerated as long as significant increase in the speed of plasmon generation is attained[13]. The developments in these two niches have been smoothly proceeding along two parallel tracks, but now and then the claims of "quantum phenomena" have made their way into the description of hot carriers' action [14, 15]. Since "quantum effects" are rather loosely defined, these claims can always be admissible, as every known phenomenon in the microscopic world can indeed be accurately described using the laws of quantum theory. But, from practical point of view, it is far more important to address an entirely different question of whether invoking the laws of quantum physics improves one's understanding of hot carrier phenomena? What is even more crucial is to determine when and where these phenomena cannot be adequately described classically and quantum theory becomes indispensable? This work is an attempt to answer the last question.

As of today, there already exists an imposing body of work [16-18] in which quantum theory (ranging from particle in the box [19] approach all the way to density functional [20]) has been applied to surface plasmon polaritons (SPPs) on nanoparticles of various shapes and sizes. Practically all these theoretical works [21, 22] as well as experimental data [23, 24] indicate that while quantum corrections do matter, they do not change the picture qualitatively and for the most part can be taken into account by introducing size-dependent damping constant [23, 25]. There is little need to replicate these calculations. The present effort, not being a review article, is intended to neither pay tribute to all the prior works nor to criticize them, and the reader who is primarily looking to find out whether a particular work has been given a proper accolade, may be disappointed. The intent here is to use the briefest, simplest and most transparent of arguments to explain to a broad audience why the impact of quantum effects on hot carrier generation have been so far found to be quite limited, and then to lead that audience towards the discovery that quantum effects do play a very important role, but not exactly where they habitually have been looked for.

With that in mind, we consider first the decay of SPPs in a metal nanoparticle that produces hot carriers. While it is expected that as the size of nanoparticle decreases, the quantum effects must come into play sooner or later we show that the quantum effects become significant only when the size of a nanoparticle $d$ approaches the value of $d_{\min} = v_F / f$ where $v_F$ is the Fermi velocity and $f$ is optical frequency. For visible light it amounts to only 2-3 nanometers. Up to that limit light absorption and SPP decay can be adequately described by what essentially is a Drude formula with a simple continuous $1/f^2$ frequency and $1/d$ size dependencies of damping constant. At the same time, when it comes to statistics of SPPs and carriers engendered by the latter's decay, unexpectedly, quantum mechanics must be always invoked since even nanoparticles as large as 200nm under realistic illumination never have more than a single SPP residing on each of them (and most often no SPP at all). That leads to interesting (and in a sense surprising) observation that instant increase of the electron temperature in nanoparticles depends only on their volumes and not on the intensity of the illumination (within wide limits). Moreover, under continuous illumination the instant ("quantized") increase of electron temperature in smaller nanoparticles can be orders of magnitude higher than average electron temperature rise evaluated classically, with important consequences for various hot carrier applications. Below, the basis for these conclusions are given in greater detail.

**When does metal loss become quantized?**

Absorption of light in metals at long wavelengths, for which the photon energy is below the interband transition threshold (e.g. 350nm for Ag and 500 nm for Au) is due to intraband transitions of free carriers that are forbidden by the momentum conservation rules. The reason metals absorb at all is that momentum conservation is maintained with the assistance of electron scattering on phonons, defects and impurities as well as on other electrons[3]. The combined rate of momentum damping in the bulk metal is $\gamma_{bulk}$ (typically on the scale of $10^{14}s^{-1}$) and phenomenologically it can be introduced into Drude formula for the dielectric constant $\varepsilon(\omega) = \varepsilon_\infty - \omega_p^2 / (\omega^2 + i\omega\gamma)$ where $\omega_p$ is the plasma frequency. Then the absorption rate in the metal can be evaluated as $R_{abs}(\omega) = \omega \text{Im}(\varepsilon) = (\omega_p^2/\omega^2)\gamma_{bulk}$ If an SPP is excited on the metal surface or on a nanoparticle, the fraction of electric field energy in the metal is always $\omega^2/\omega_p^2$ and therefore $\gamma_{bulk}$ is precisely the rate of surface plasmon decay. This statement also directly follows from the fact that in the absence of magnetic field (which is a fact in sub-wavelength nanoparticles), half of the time the entire energy of SPP is stored in the metal as kinetic energy of electrons which decays with twice momentum relaxation rate $\gamma_{bulk}$.

Phonons and imperfections, however, are not the only pathways enabling intraband transitions as electron reflection from the sharp metal boundary can also provide momentum matching – this is the basis of the so-called Kreibig [26, 27] decay channel in nanoparticles that can be phenomenologically quantified by augmenting the bulk scattering by the additional scattering on the surface at a rate $\gamma_{surf} \sim v_F / d$ equal to the frequency of collisions with the surface. A more detailed treatment [28, 29] shows that momentum-dependent (or nonlocal) dielectric function of metal $\varepsilon(k)$ [30] acquires non-zero imaginary part for wave vectors $k > \omega/v_F$, where $v_F$ is the Fermi velocity, hence for large wavevector direct (unassisted) intraband transitions damping become possible. This mechanism is called Landau damping (LD). In the presence of a sharp interface, the electric field of a plasmon does contains such large wavevector components, which leads to the additional surface-induced, or LD rate [31] $\gamma_{LD} = \frac{3}{8} v_F / d_{eff}$, where $d_{eff}$ is the effective "depth" of the field which is essentially the volume-to-surface ratio

$$d_{eff} = \int_{metal} E(\mathbf{r})^2 dV / \int_{metal\ surface} E_\perp^2(\mathbf{r}) dS \qquad (1)$$

and $E_\perp$ is the normal to the surface component of the electric field inside the metal. Eq. 1 leads to $\gamma_{LD} = \frac{3}{4} v_F / d$ for both sphere of dimeter d and cube with a side d, which is very close to the result phenomenologically derived by Kreibig [26, 27].

As one can see, LD becomes the dominant damping mechanism once $d < v_F / \gamma_{bulk} \sim L_{mfp}$, $L_{mfp}$ being the mean free path of an electron within the metal. At this point, however, one should expect the energy levels inside the nanoparticle to become quantized, with a discreet set of eigen-energies and transitions between them. This is what after all occurs in semiconductor nanoparticles-quantum dots (QDs) [32-34] where quantization has been exploited in numerous practical devices such as lasers and photodetectors. This raises a question of whether similar quantization effects can be observed in metal nanoparticles. Furthermore, the LD estimate is based on nonlocality of the dielectric function (which can be expressed as either $\varepsilon(\mathbf{k})$ or $\varepsilon(\mathbf{r}_1, \mathbf{r}_2)$ and for the quantized structure the whole concept of dielectric constant makes no sense and one must account for the polarizability of the entire particle instead. And this poses a question: what is the impact of quantum confinement on the intraband absorption and consequently on the SPP damping, and when does the latter starts deviating significantly from the one predicted by Kreibig?

We now look into the decay of a uniform electric field $E$ inside a cubic nanoparticle and more specifically how it depends on the size of the nanoparticle $d$ and on the oscillating frequency of the field $\omega$. The confined wavefunctions are $\psi_{mpq} = \psi_m(x)\psi_p(y)\psi_q(z)$, where $\psi_m(x) = \sqrt{2/d} \sin(m\pi x/d)$. Since all the energies are close to Fermi level, the condition $m^2 + p^2 + q^2 \approx m_0^2 v_F^2 d^2 / \pi^2 \hbar^2$ is maintained and the indices run from 1 to $N \approx dm_0 v_F / \pi\hbar$. The relation between the index n along a given direction x and Fermi velocity projection on that direction is $v_{F,x} \approx n\pi\hbar / dm_0$. Then the matrix element of Hamiltonian for x-polarized light is

$$H_{mn} = \frac{1}{2} eE \int \psi_{mpq} x \psi_{npq} dxdydz = 4eEd \frac{mn}{\pi^2(m^2 - n^2)^2}, \qquad (2)$$

as long as *m* and *n* have different parities. Furthermore, since the transition energy is $\hbar\omega = \pi^2\hbar^2(m^2-n^2)/2m_0d^2$ one can transform (2) into $H_{mn} = eEv_{F,x}^2/d\omega^2$ and using the expression for the energy density inside the metal $U = \varepsilon_0 E^2 d^3/2$ we obtain:

$$H_{mn}^2 = 2\frac{e^2}{\varepsilon_0 d}\frac{v_{F,x}^4}{d^4\omega^4}U \tag{3}$$

Since the separation between two nearest odd or nearest even states is $\Delta E = \pi^2\hbar^2[n^2-(n-2)^2]/2m_0d^2 \approx 2\pi\hbar v_{F,x}/d$, one can introduce a 1D density of states per unit energy in the entire cube as $\rho_x = 1/\Delta E$. Next we can apply Fermi Golden rule to evaluate the rate of energy loss by the field inside the cube as

$$R_{abs} = \frac{dU}{dt} = -\hbar\omega \times \frac{2\pi}{\hbar}\langle H_{mn}^2 \rho_x \rangle_{mpq} \times \hbar\omega\rho(E_F)d^3 = -2\frac{e^2}{\varepsilon_0 d}\frac{\langle |v_{F,x}^3| \rangle}{\omega^2}\rho(E_F)U, \tag{4}$$

where $\hbar\omega\rho(E_F)d^3$ is the total number of sates within $\hbar\omega$ from the Fermi level and the averaging is done over the indices $m, p, q$ i.e. over the angle $\theta$ so that $\langle |v_{F,x}^3|\rangle = v_F^3\langle |\cos^3\theta|\rangle = v_F^3/4$. For spherical conduction band one has the following relation between the plasma frequency and properties at Fermi level $\omega_P^2 = e^2 v_F^2 \rho(E_F)/3\varepsilon_0$ which brings us to $R_{abs} = -(3/2)(\omega_P^2/\omega^2)(v_F/d)$, therefore the damping rate of plasmon due to confinement is $\gamma_{conf} = \tfrac{3}{2} v_F/d$.

Interestingly, while the confinement damping follows essentially the same dependence as LD, its value for the cube is twice as high. For this there is a simple explanation. Using LD theory one assumes that the reflections on two opposite sides of the cube that enable intraband transitions occur independent of each other so the squares of the transition amplitudes add up. But if the electron experiences no collisions between the two walls, then the amplitudes of transition should add up *coherently*, rather than the incoherently, hence the factor of two is due to quantum interference. Of course, since the scattering is a statistical process, the damping rate changes gradually as the size decreases, $\gamma_{conf} = \tfrac{3}{4}[1+\exp(-d/L_{mfp})]v_F/d$. But what is most relevant to our discussion is that the expression for damping does not contain Planck's constant, hence it can be perfectly well described classically.

To verify these simple estimations, we perform a numerical modeling of the decay process where we plot the decay rate of energy inside the metal $R_{abs}(\omega)$ as a function of frequency for different cube diameters in Fig1.a. The rate is calculated by adding up oscillator strengths of all the transitions and then broadening them by $\Gamma \sim 60 meV$ to account for the bulk scattering. As one can see, the decay rate decreases as $\omega^{-2}$, just as expected from analytical calculations, for all but the smallest (d<4nm) cube sizes. Also one can see the expected increase of $R_{abs}(\omega)$ with the decrease of size. Note that $R_{abs}(\omega)$ simply describes the loss of energy in the metal and not the absorption of incident radiation by the nanoparticles – hence the absence of resonances. Shown in Fig. 1b is the plasmon damping rate due to confinement $\gamma_{conf}(\omega) = R_{abs}(\omega)\omega^2/\omega_P^2$ - the rate is nearly frequency independent for $d \geq 4nm$, as one would expect from the most simple Kreibig's [26, 27] theory as well as from more detailed LD decay theory based on Lindhard's formula[29]. Deviations occur only for smaller cube sizes – one can attribute the decreased absorption at low energies for d=2nm to quantization of energy levels in the vicinity of Fermi level and appearance of energy gap. Finally, in Fig.1c we plot the plasmon decay rate averaged over the range of frequencies $\langle \gamma_{conf}(\omega) \rangle_\omega$ versus the cube size d – as one can see one obtains a nearly perfect 1/d dependence with numerical value close to $\tfrac{3}{2} v_F / d$.

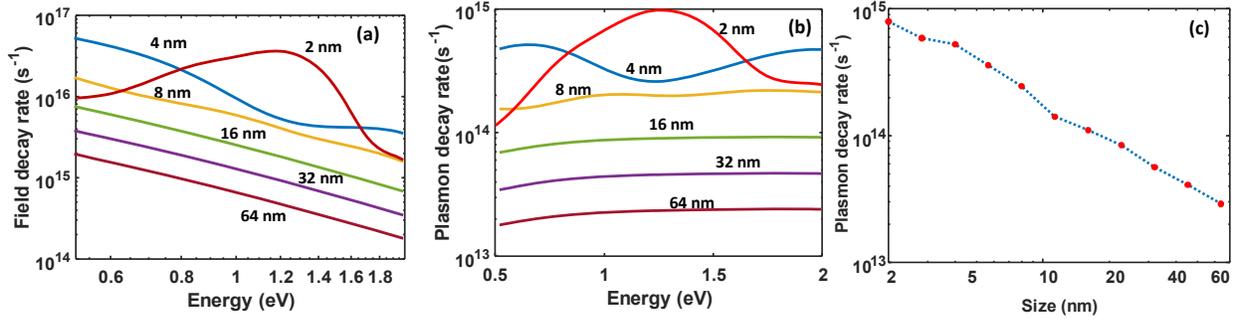

**Figure 1** (a) rate of loss (absorption) for the energy confined within the metal cube $R_{abs}(\omega)$ as a function of frequency for different cube sizes. The energy loss rate is inversely proportional to the square of frequency. (b) SPP polariton decay rate due to confinement $\gamma_{conf}$ vs frequency for different cube sizes. This rate is nearly frequency (wavelength) independent. (c) SPP polariton decay rate due to confinement averaged over photon energies from 0.5 to 2eV vs cube size. The dependence is roughly 1/d in agreement with the phenomenological Kreibig's theory.

According to this simple exercise, the quantization becomes important when the smallest transition energy between two levels in the vicinity of Fermi level $2\pi\hbar v_F / d\sqrt{3}$ (factor of $\sqrt{3}$ comes from averaging over all directions) exceeds photon energy $\hbar\omega$ which immediately leads do the condition $d < (2\pi/\sqrt{3})v_F/\omega$. Phenomenologically, this condition simply means that the electron reaches the surface within less than a single optical period, so the "free electron" description is no longer valid. It can be just as easily interpreted from Lindhardt's theory, if one notices that it corresponds to the situation when the breadth of spatial spectrum of the electric field $\Delta k \sim \pi d^{-1}$ exceeds the offset of the LD in Lindhard's formula, $\omega/v_F$ which makes the approximations made in [28, 29] invalid. For the ~1-1.5eV energy range this corresponds to about 2-3 nm and it is easy to see from Fig.1 that it is there that the curves start to deviate from simple estimate and show an oscillatory pattern. Therefore, when it comes to the effect of size reduction on the absorption by nanostructures, for as long as dimensions exceed a couple of nanometers, the "quantum effects" are quite adequately taken into account by the broadening of the SPP resonance also accompanied by relatively small frequency shift. At smaller dimensions a more discrete spectrum emerges, but given how broad the resonances are, that plays very little role, especially since in many practical applications such as photovoltaics and photo catalysis the incoming light is broadband. Calculations involving full quantum mechanical model [15, 19, 35-37] lead to essentially the same limits of applicability of classical expressions, but we have arrived at this conclusion without resorting to any complicated quantum mechanical calculations, by rather straightforward and physically transparent reasoning. Clearly, the striking distinctions between metal nanoparticles and QD's can be also explained simply by the fact that while a few nm in size QD is typically occupied by only a few electrons, in metal nanoparticles of similar sizes the number of electrons is measured by hundreds and thousands, so the collective modes of electrons in metal bear very little resemblance to the single electron transitions in quantized semiconductor structures as first noted in [38].

**When do the SPPs and hot carriers start following quantum rules?**

Having not discovered any significant quantum effects when it comes to description of the larger electrons whose collective motion comprises SPP it now makes sense to discuss whether there is any quantization when it comes to the hot carriers engendered by the decay of SPP. To do so, let us find out the number of SPPs that reside at a given nanoparticle at a given time. One way to find out this number is to introduce the absorption cross-section by a metal nanoparticle of volume $V$, $\sigma_{abs} \sim AV\omega^2/\gamma c$, where $A$ is shape dependent parameter of the order of 1 (for spherical nanoparticle A=3) and $\gamma$ is the total SPP damping rate. Therefore, if the incoming light power density (irradiance) is $I_{in}$ then the average number of SPPs on a nanoparticle is $N_{SPP} = I_{in}\sigma_{abs}\gamma^{-1}/\hbar\omega \sim AV(\omega/\gamma)^2 I_{in}/\hbar\omega c$. This is easy to interpret as electric field in the SPP (or any oscillator for that matter) gets enhanced by $Q \sim \omega/\gamma$. In noble metals Q rarely exceeds 10 in Au and 15 -20 in Ag, especially for smaller particles subject to LD. Therefore for visible light we can estimate the irradiance required to sustain a single SPP on a nanoparticle is roughly $I_1 \sim 10^{10}/V$ where the volume is in nm$^3$ and irradiance is in W/cm$^2$. Then for a 50 nm nanoparticle a 100kW/cm$^2$ input power density is required to sustain a single SPP in the mode and for a 10nm nanoparticle the value is as high as 10MW/cm$^2$.

Now, these power densities are routine in the femtosecond and picosecond lasers used to study dynamics of hot carriers in numerous works [39-42]. Then multiple SPPs can be excited on each nanoparticle which in their turn engender large number of hot carriers that then decay towards equilibrium. However, these enormous power densities have very little to do with the real world scenarios in which hot carrier based technologies are expected to play a role. If we consider applications such as photo catalysis[43-45] or photovoltaics[46], then the light source is the sun with irradiance of only about 0.1W/cm$^2$ so even if one uses a powerful concentrator with 1,000 concentration ratio, the average number of SPPs on a 10-50 nm nanoparticle $N_{SPP}$ is anywhere between 10$^{-5}$ and 10$^{-3}$. Likewise, if one considers photodetectors [6, 47], the typical power density arrives at the detector plane is typically way below a kW/cm$^2$. Thus, even if an optical antenna[48] is employed to further concentrate the energy, no more than a single SPP ever populates a given nanoparticle at a given time. To understand what it all means, consider Fig.2a in which a rendering of nanoparticle array "frozen" at some time t$_0$ is shown with only one out of $N_{SPP}^{-1} \gg 1$ nanoparticles having an SPP excited on it.

This perhaps unforeseen outcome is simply the consequence of a very short SPP lifetime $\tau_{spp} = \gamma^{-1} \sim 10 fs$. Even putting aside the limited incident power considerations, it is simply impossible to sustain even a single SPP on a nanoparticle for a long time before it melts, the argument first made in reference to "spaser" in [49]. If $N_{SPP} = 1$ is to be maintained, the power dissipated per nanoparticle would have to be in the order of $P_1 = \gamma \hbar \omega \sim 20 \mu W$ amounting up to $10^{11} W/cm^3$ power density in 50nm nanoparticle. That would cause a drastic temperature rise of about 1000K in ~10ns, no matter how efficient is the cooling, with dire consequences of immediate meltdown of the nanoparticle. So, clearly the process of SPP generation and decay is discrete, or quantum (as evidenced from the presence of Planck's constant in the expression for N$_{SPP}$) and this fact has important repercussions.

First of all, one cannot use the term "dephasing" [50] when looking at SPPs at practical excitation power densities – a single SPP cannot de-phase simply because it has no phase reference, i.e. there is no second SPP on the same nanoparticle. Instead, it can only be annihilated. Second, SPP cannot gradually lose its energy to the movement of electron-hole pairs near the Fermi surface as claimed in a number of works where the "Drude" [51, 52]or "friction" [19, 35] like mechanisms are somehow introduced. That would violate the energy conservation or it would require the SPP to change frequency! The SPP decay is quantum and once SPP is annihilated a single electron hole pair (or two of them if electron-electron scattering causes SPP decay) is created with entire (or nearly entire if a phonon has been involved) energy is now contained in these hot carriers. It means that the discrete or quantum features have been also transferred to the hot carriers and at a given time only a relatively few hot carriers are present.

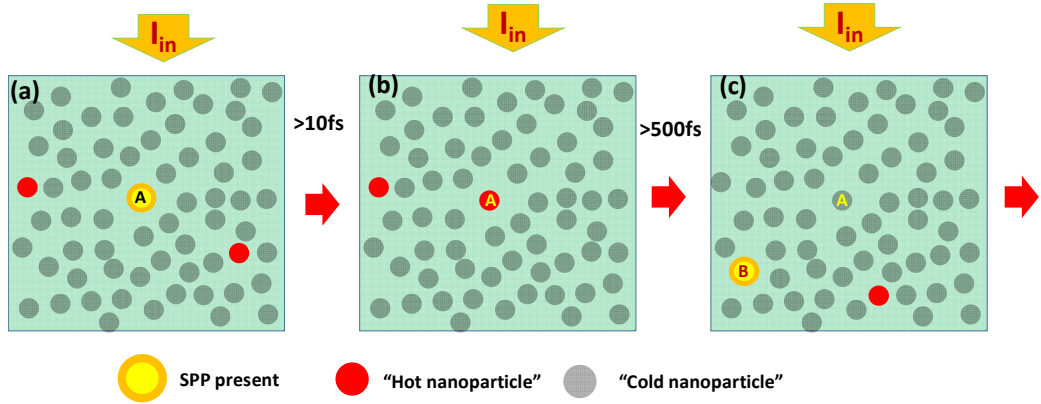

**Figure 2** Evolution of SSPs and hot carriers in the ensemble of nanoparticles. (a) An SPP is generated on one out of great many nanoparticles (b) SPP has decayed and the electron temperature $T_e$ of the nanoparticle A has risen above lattice temperature T (c) The electron temperature of nanoparticle A has reached equilibrium with the lattice.

Once the "primary" electron and hole have been generated, they thermalize due to rapid electron-electron collisions that occur on the rate of tens of fs [42, 53-55]. Since in each scattering act the energy of hot carrier is divided between three new, "secondary" carriers, it only takes a few collisions to thermalize the electrons to the electron temperature exceeding the lattice temperature by $\Delta T_e$ [56]. This elevated temperature will then persist for some time $\tau_{EL}$ that is on a scale of hundreds of femtoseconds because that is how long it takes for the electron-phonon scattering to restore equilibrium with the lattice. Note that $\tau_{EL}$ is so long not because the electron-phonon scattering rate is weak (it is not! [57]) but because the energy transferred to lattice each time phonon scattering takes place is only a phonon energy. Therefore the fraction of nanoparticles that have elevated electron temperature at a given time is $N_{hot} = N_{SPP} \gamma \tau_{EL}$, i.e. 10-100 times $N_{SPP}$. For the realistic irradiances that are less than a 10kW/cm² $N_{hot} \ll 1$ and from this fact a rather important consequence for practical applications follows.

As shown in Fig. 2b – a freeze frame taken after some time $t_1 = t_0 + \Delta \tau_1$ where $\tau_{SPP} < \Delta \tau_1 < \tau_{EL}$ the original nanoparticle "A" that had SPP at time $t_0$ remains "hot", i.e. has elevated electron temperature $T_e$. A few other nanoparticles excited by SPP's within time interval from $t_1 - \tau_{EL}$ until $t_1$ are also hot. At time $t_2 > t_0 + \tau_{EL}$ the nanoparticle "A" has cooled down to the lattice temperature T, as shown in Fig. 2c, yet there is a chance that an SPP got excited on another nanoparticle "B", and the sequence of events repeats itself.

The value of $\Delta T_e = T_e - T$ can be easily found as $\Delta T_e = \hbar\omega/c_{el}V$ where the electron specific heat is $c_{el} = \pi^2 k_B^2 T N_e / 2E_F$ [58], $N_e \sim 60$ $nm^{-3}$ is the electron density and $E_F$ is the Fermi energy ($E_F$ = 5.5eV for both Au and Ag). The rise in temperature does not depend on the incident power and is determined solely by photon energy and the volume – if this is not a quantum effect, we are indeed at a loss to say what is? We then obtain $\Delta T_e \sim 1.6\times 10^4 / V$ $(K)$ where the volume is in nanometers cube. In large nanoparticles (e.g. d>25nm), the temperature rises by less than 2K. Yet, for smaller nanoparticles, say 5nm length the temperature rise can be by as high as 200K, and for 3nm nanoparticle it exceeds 1000K.

In the discretized (or quantum) picture the electron temperature is nearly instantly elevated by $\Delta T_e$ for the time $\tau_{EL}$ after each SPP generating event while these events occur with a frequency $\gamma N_{SPP}$ i.e. $N_{hot}$ is a duty cycle. Therefore, the average rise of electron temperature, the one which could be calculated with disregard the quantum nature of the process is $\Delta \bar{T}_e = \Delta T_e N_{hot} = A(\omega/\gamma)^2 I_{in} \gamma \tau_{EL} / cc_{el}$, a pure classical result that depends on neither photon energy nor the particle size. The average, or classical rise of electron temperature is much less than an instant one [43, 59], but it takes place continuously and not in short bursts. This can be seen from Fig.3 where both discrete or instant rise in electron temperature $\Delta T_e$ (solid line) and the average rise $\Delta \bar{T}_e$ for three different input irradiances (dashed lines) are plotted versus the nanoparticle size

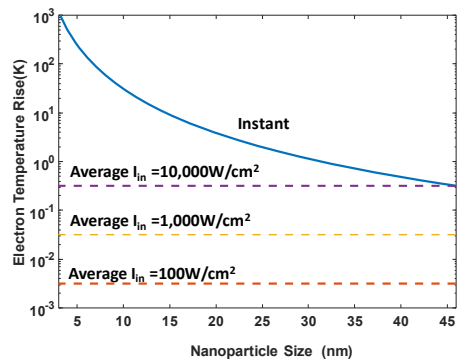

**Figure 3.** Instant (solid line) and Average (dashed lines) electron temperature rise vs the size of the nanoparticle

Now, let us see what difference does "quantization" make? We posit that the desired hot carrier induced reaction can be only undertaken by a carrier with an energy higher than a certain barrier energy $\Phi$ taken relative to the Fermi energy. Then, when the carriers are at equilibrium with the lattice the rate at which the desired action occurs is $R_0 = \exp(-\Phi/k_B T)$ and when the average temperature is elevated to $\overline{T}_e = T + \Delta\overline{T}_e$ with a duty cycle $N_{hot}$ the time-averaged or "classical" reaction rate becomes $R_{av} = \exp(-\Phi/k_B \overline{T}_e) \approx R_0 R_0^{-\frac{\Delta\overline{T}_e}{T}}$ and the enhancement $K_{av} = R_{av}/R_0 = R_0^{-\frac{\Delta\overline{T}_e}{T}}$ When the granularity is taken into account, the instant rise in electron temperature is $\Delta T_e = \Delta\overline{T}_e / N_{hot}$ but it occurs with a low duty cycle $N_{hot}$, so that the "discrete" or "quantum" enhancement is

$$K_{dis} = (1 - N_{hot}) + N_{hot} \exp(-\Phi/k_B T_e)/R_0 \approx 1 + (K_{av}^{1/N_{hot}} - 1)N_{hot} \geq K_{av} \quad (5)$$

In Fig.4a the dependence of the reaction rate enhancement $K_{dis} - 1$ on heating duty cycle $N_{hot}$ for different values of $K_{av}$ is shown. The horizontal line corresponds to a 100% enhancement. For relatively large duty cycles $N_{hot} > 0.01$ the lines are horizontal, i.e. quantization does not cause any change in the estimate of the reaction rate. But for small duty cycle (i.e. for small nanoparticles) the enhancement increases rapidly and for really small duty cycles the difference made by quantization becomes enormous. To put it into a different reference frame we plot the enhancement vs nanoparticle size for different barrier heights in Fig.4 b and c for two different incident irradiances. As one can see, for relatively high barriers and small nanoparticles the difference between the quantized and classical results is indeed enormous.

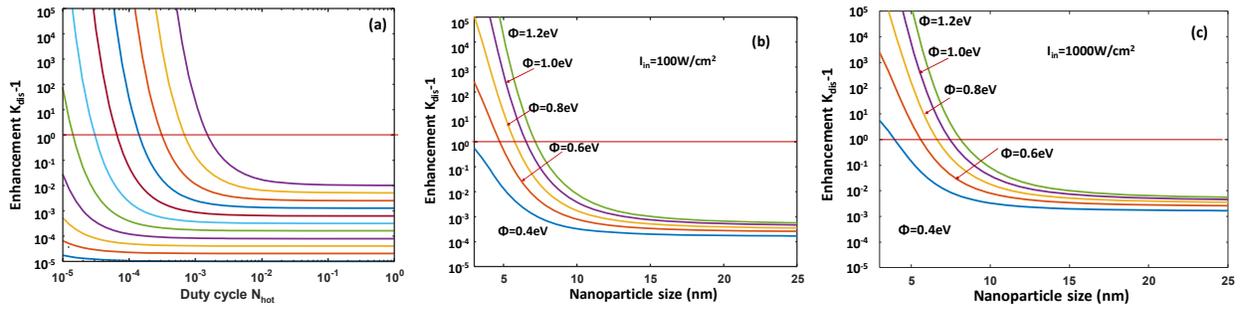

**Figure 4** (a) Enhancement of the hot electron induced reaction as a function of heating duty cycle (the fraction of time the nanoparticle contains hot carriers) (b,c) Enhancement of the hot carrier induced reaction vs. the size of nanoparticle for two different incident irradiances

The discrete character of hot carrier generation is manifested in significant enhancement of desired hot carrier action in comparison to classical one. Therefore, while the classical theory with pitifully small rise in electron temperature cannot possibly explain how the rate of thermionic processes can be enhanced by so much [43, 59], the quantum theory offers a very plausible explanation. Of course, in case when the barrier is very high and $R_0$ is small, neither classical nor quantum theory offer any explanation. It can be the action of ballistic (or the first generation) hot carriers that have not yet experienced a single collision [56] , which is precisely what happens in Schottky photodetectors[6, 47]. Or it can be due to rise in the lattice temperature, i.e. not be a hot carrier effect in the first place[59].

**Conclusions**

The lesson that may be drawn out of this unassuming exercise is rather short and plain, but hopefully not inconsequential. The sheer fact that one can employ full quantum narrative to precisely describe a given process does not necessarily mean that doing so will reveal anything that could not be adequately described using rudimentary classical approach. This platitude can be applied to any task, and this study of hot carrier processes in plasmonic nanoparticles is no different. When it comes to depiction of metal nanoparticles, the metal is perfectly well described classically using a Drude model with damping constant modified by surface collisions for all but the smallest (less than 2-3nm) nanoparticles. Even for smaller nanoparticles quantum correction amounts to very little if the incoming radiation is broadband. At the same time, granular, or quantum character of SPP excitation and decay cannot be neglected. The "spiking" or shot-noise-like nature of the rise and fall of electron temperature may lead to orders of magnitude difference in the rate of certain hot-carrier processes, such as for instance photo catalysis. With any lack, this plea for judicious use of quantum theory will be heard by the community and be of some use to it.

**Acknowledgement**

Authors acknowledge support of Air Force Office of Scientific (AFOSR) Research Award # FA9550-16-10362 -and the Israeli−USA Binational Science Foundation (BSF). JBK also appreciates indispensable assistance and critique offered by Prof. P. Noir and sharp comments by the newest post-doctoral member of his group Ms. S. Artois.

**Figure Captions**

**Figure 1** (a) rate of loss (absorption) for the energy confined within the metal cube $R_{abs}(\omega)$ as a function of frequency for different cube sizes. The energy loss rate is inversely proportional to the square of frequency. (b) SPP polariton decay rate due to confinement $\gamma_{conf}$ vs frequency for different cube sizes. This rate is nearly frequency (wavelength) independent. (c) SPP polariton decay rate due to confinement averaged over photon energies from 0.5 to 2eV vs cube size. The dependence is roughly 1/d in agreement with the phenomenological Kreibig's theory.

**Figure 2** Evolution of SSPs and hot carriers in the ensemble of nanoparticles. (a) An SPP is generated on one out of great many nanoparticles (b) SPP has decayed and the electron temperature $T_e$ of the nanoparticle A has risen above lattice temperature T (c) The electron temperature of nanoparticle A has reached equilibrium with the lattice.

**Figure 3.** Instant (solid line) and Average (dashed lines) electron temperature rise vs the size of the nanoparticle

**Figure 4** (a) Enhancement of the hot electron induced reaction as a function of heating duty cycle (the fraction of time the nanoparticle contains hot carriers) (b,c) Enhancement of the hot carrier induced reaction vs. the size of nanoparticle for two different incident irradiances